\documentclass[a4paper]{ifacconf}

\usepackage{graphicx}      
\usepackage{natbib}        
\usepackage{amsmath,amssymb,bm} 
\usepackage{acronym}
\usepackage{color}
\usepackage{graphicx}		
\usepackage{mathtools}
\usepackage[normalem]{ulem}
\usepackage{comment}
\usepackage{algorithm,algpseudocode}
\usepackage{xcolor}
\usepackage{caption}
\usepackage{subcaption}
\usepackage{tikz}
\usepackage{pgfplots}

\definecolor{mygray}{gray}{0.35}
\newcommand{\pink}[1]{\textcolor{magenta}{#1}} 

\def\IR{\mathbb R}					

\newcommand{\defeq}{\vcentcolon=}

\newcommand{\Scal}{\mathcal{S}} 
\newcommand{\I}{\mathbf{I}} 
\newcommand{\mbf}[1]{\mathbf{#1}} 
\newcommand{\bmat}[1]{\begin{bmatrix} #1 \end{bmatrix}} 
\newcommand{\Xcal}{\mathcal{X}}

\newcommand{\B}{\mathbf{B}} 
\newcommand{\x}{\mathbf{x}} 
\newcommand{\y}{\mathbf{y}} 
\newcommand{\z}{\mathbf{z}} 
\newcommand{\bP}{\mathbf{P}} 
\newcommand{\Q}{\mathbf{Q}} 

\newcommand{\X}{\mbf{X}}

\newcommand{\vecdim}[1]{{\in} \mathbb{R}^{#1}}

\newcommand{\norm}[1]{\left\lVert#1\right\rVert}
\newcommand{\absVal}[1]{\left\lvert#1\right\rvert}

\newcommand{\Pre}{\text{Pre}}
\newcommand{\Suc}{\text{Suc}}

\newcommand{\Qcal}{\mathcal{Q}}

\newcommand{\Mcal}{\mathcal{M}}
\newcommand{\Kcal}{\mathcal{K}}
\newcommand{\Ncal}{\mathcal{N}}
\newcommand{\Bcal}{\mathcal{B}}
\newcommand{\sumLam}{\sum_{k=0}^m\lambda_k}
\newcommand{\Ecal}{\mathcal{E}}
\newcommand{\Dcal}{\mathcal{D}}

\newcommand{\thmref}{Theorem }
\newcommand{\lemref}{Lemma }
\newcommand{\defref}{Definition }
\newcommand{\asmptref}{Assumption }
\newcommand{\assumeref}[1]{Assumption~\ref{#1}}
\newcommand{\figref}{Figure }
\newcommand{\conref}{Constraint }
\newcommand{\condref}{Condition }
\newcommand{\alref}{Algorithm }
\newcommand{\algoref}[1]{Algorithm~\ref{#1}}
\newcommand{\secref}{Section }

\newcommand{\lemrefs}{Lemmas }
\newcommand{\defrefs}{Definitions }
\newcommand{\asmptrefs}{Assumptions }

\newcommand{\conrefs}{Constraints }
\newcommand{\condrefs}{Conditions }

\acrodef{dl}[{DL}]{deep learning}
\acrodef{rl}[{RL}]{reinforcement learning}
\acrodef{nn}[{NN}]{neural network}
\acrodef{dnn}[{DNN}]{deep neural network}
\acrodef{tdl}[{TDL}]{temporal difference learning}
\acrodef{pid}[{PID}]{proportional–integral–derivative}
\acrodef{us}[{US}]{Ultrasound}
\acrodef{mse}[{MSE}]{mean squared error}
\acrodef{sgd}[{SGD}]{stochastic gradient descent}
\acrodef{ico}[{ICO}]{iterative convex overbounding}
\acrodef{lmi}[{LMI}]{linear matrix inequality}
\acrodef{mjls}[{MJLS}]{Markov jump linear system}
\acrodef{io}[{IO}]{input-output}
\acrodef{iqc}[{IQC}]{integral quadratic constraints}
\acrodef{cnn}[{CNN}]{convolutional neural network}
\acrodef{il}[{IL}]{Imitation learning}
\acrodef{mpc}[{MPC}]{model predictive control}
\acrodef{sdp}[{SDP}]{semi-definite programming}
\acrodef{relu}[{ReLU}]{rectified linear unit}
\acrodef{us}[US]{ultrasound}
\acrodef{mdp}[MDP]{Markov Decision Process}
\acrodef{iid}[iid]{identical and independently distributed random variable}
\acrodef{pid}[PID]{Proportional Integral Derivative}

\acrodef{lqr}[LQR]{linear-quadratic regulator}

\acrodef{cpa}[CPA]{continuous piecewise affine}
\acrodef{hji}[HJI]{Hamilton Jacobi Inequality}
\acrodef{roa}[ROA]{region of attraction}
\acrodef{pi}[PI]{positive invariant}
\acrodef{uub}[UUB]{uniform ultimate boundedness}

\acrodef{gaio}[GAIO]{Global Analysis of Invariant Objects}

\newcommand\copyrighttext{%
  \footnotesize \textcopyright 2025 the authors. This work has been accepted to IFAC for publication under a Creative Commons Licence CC-BY-NC-ND.}
\newcommand\copyrightnotice{%
\begin{tikzpicture}[remember picture,overlay]
\node[anchor=south,yshift=60pt] at (current page.south) {\fbox{\parbox{\dimexpr\textwidth-\fboxsep-\fboxrule\relax}{\copyrighttext}}};
\end{tikzpicture}%
}

\begin{document}
\begin{frontmatter}

\title{Data-driven certificates of constraint enforcement and stability for unmodeled, discrete dynamical systems using tree data structures\thanksref{footnoteinfo}}

\thanks[footnoteinfo]{This work was supported by NSF Grant CMMI-2303157. Any opinions, findings, and conclusions or recommendations expressed in this material are those of the authors and do not necessarily reflect the views of the NSF.}

\author[First]{Amy K. Strong} 
\author[Second]{Ali Kashani} 
\author[Second]{Claus Danielson}
\author[First]{Leila J. Bridgeman}

\address[First]{Duke University, Durham, NC, 27708, USA (e-mail: amy.k.strong@duke.edu).}
\address[Second]{University of New Mexico, Albuquerque, NM, 87131, USA}

\begin{abstract}                
This paper addresses the critical challenge of developing data-driven certificates for the stability and safety of unmodeled dynamical systems by leveraging a tree data structure and an upper bound of the system's Lipschitz constant.
Previously, an invariant set was synthesized by iteratively expanding an initial invariant set. In contrast, this work iteratively prunes the constraint set to synthesize an invariant set -- eliminating the need for a known, initial invariant set. Furthermore, we provide stability assurances by characterizing the asymptotic stability of the system relative to an invariant approximation of the minimal positive invariant set through synthesis of a discontinuous piecewise affine Lyapunov function over the computed invariant set. 
The proposed method takes inspiration from subdivision techniques and requires no prior system knowledge beyond Lipschitz continuity. 
\end{abstract}


\end{frontmatter}

\section{Introduction}
Establishing the safety and stability of unmodeled systems is challenging.
\copyrightnotice

Invariant sets and Lyapunov functions are related. Invariant sets are regions of the state space wherein a dynamical system remains for all time, which is vital for constraint adherence.
Lyapunov functions are associated with the energy of a dynamical system and can be used to verify stability of the system about an equilibrium (\cite{lasalle2012stability}) or about a set (\cite{blanchini2008setTheo}) -- often the minimum \ac{pi} set within the constraint set.
For nonlinear systems, stability is often a local property and thus requires knowledge of the \ac{roa}, the region in which the system is Lyapunov stable with respect to an equilibrium point or destination set. 
The \ac{roa} depends on the specific Lyapunov function constructed, and asymptotic stability within a \ac{roa} does not necessarily imply constraint satisfaction.
Therefore, the \ac{roa} may not coincide with the maximal \ac{pi} set, the largest invariant set within the system constraints, leading to a disconnect between stability and safety guarantees.

For nonlinear systems, invariant sets and Lyapunov functions are difficult to construct even if the model is known. 
Methods to determine an invariant set are often not guaranteed to converge for nonlinear systems (\cite{kerrigan2000robust}), only give approximate invariant sets, (\cite{dellnitz2002set};\cite{korda2014controller}), and/or are restricted to certain classes of systems (\cite{korda2014controller}). 
Invariant sets can also be synthesized as sublevel sets of a Lyapunov function (\cite{lasalle2012stability}), but the Lyapunov function of a stable nonlinear system has no set form and is only valid for a potentially unknown \ac{roa}. Model based methods to determine Lyapunov functions often choose a flexible function form and develop a constrained optimization problem to synthesize the Lyapunov function (\cite{giesl2014computation, anderson2015advances}), an approach leveraged in this paper.


This paper presents a deterministic, data driven and model free approach to synthesize invariant sets and Lyapunov functions. While methods exist to identify system models from data, a model-free approach circumvents issues introduced in model-based methods. However, without knowledge of system behavior between sampled points, data alone is insufficient to characterize a nonlinear system. While approximations or probabilistic guarantees are an option for data driven invariant sets (\cite{korda2020computing,kashani2024probabilistic})  and Lyapunov functions (\cite{dawson2023safe}), we opt to require additional knowledge of system evolution to achieve deterministic guarantees. In line current literature (\cite{dawson2023safe}), we leverage the Lipschitz continuity.

This paper's contributions are twofold. First, we develop a data driven method to synthesize an invariant set for an unmodeled, Lipschitz continuous, discrete-time dynamical system. Our method uses deterministic sampling of successive state pairs, i.e. $\{\x,\x^+\}$, to iteratively refine a partition of the state constraint set via a tree data structure. 
We are inspired by \ac{gaio} (\cite{dellnitz2001algorithms}), a subdivision algorithm that iteratively approximates the invariant set of a known system using set-based methods. Here, we modify \ac{gaio} for the data driven case. In contrast to \ac{gaio}, our modified algorithm leverages Lipschitz continuity to produce a provably invariant set in a finite number of iterations and data points. 

The second contribution is sufficient conditions for a piecewise constant Lyapunov function within the invariant set to certify asymptotic stability of the system with respect to an invariant approximation of the minimal \ac{pi} set  -- demonstrating the \ac{uub} of the system.
These conditions are used in an optimization problem to synthesize a Lyapunov function, leveraging the same data-set and tree structure used for invariant set synthesis by assigning a Lyapunov function value to each partition of the invariant set. 

\textit{Notation and Preliminaries:}
Let $\mathbb{Z}_a^b$ be the set of integers between $a$ and $b$ inclusive.  
The interior, boundary, and closure of the set $\Omega {\subset} \mathbb{R}^n$ are denoted as $\Omega^\circ$, $\partial \Omega$ and $\bar{\Omega}$. Scalars, vectors, and matrices are denoted as $x,$ $\x,$ and $\X$. Let $\{1, -1\}^n$ denote the set of all $n$-dimensional vectors with entries either $1$ or $-1$.
Let $\varphi:\mathbb{R}_0^\infty {\rightarrow} \mathbb{R}_0^\infty$ be a class $\Kcal$ function if it is continuous, strictly increasing, and $\varphi(0){ =} 0.$ 

Define a norm ball a $\x \vecdim{n}$ for some norm, $\norm{\cdot}_p$, as $B_{r,p}(\x){\defeq} \{\y \vecdim{n} {\mid} \norm{\x-\y}_p{\leq} r\}$. Let $B_{r,\infty}(\x)$ denote a max-norm ball, where $\norm{\x}_\infty {= }\max_{i \in \mathbb{Z}_1^n} \absVal{x_i}$. Recall $\norm{\x}_\infty {\leq} \norm{\x}_2 {\leq} \sqrt{n}\norm{\x}_\infty.$ The mapping $T{:}\Xcal {\rightarrow} \mathcal{Y}$ between two metric spaces is Lipschitz continuous mapping with respect to norm $\norm{\cdot}$ if there exists some Lipschitz constant, $L {>} 0$, such that $\norm{T(\mathbf{p}) {-} T(\mathbf{q})} {\leq} L \norm{\mathbf{p} {-} \mathbf{q}}$ for all $\mathbf{p}, \mathbf{q} \in \Xcal$ (\cite{fitzpatrick2009advanced}). Define mapping $T$, and let $T^k$ indicate the mapping is applied $k$ times ($k{\in}\mathbb{Z}_{0}^{\infty}$). Let $\Delta(\x,\Scal) {\defeq} \inf_{\mathbf{s} \in\Scal} \norm{\x{-}\mathbf{s}}_2.$

\section{Problem Statement}
Consider an unmodeled discrete-time, dynamical system
\begin{flalign}\label{eq:dynSys}
    \x^+ = T(\x), \quad \x \in \Xcal \subset \mathbb{R}^n,
\end{flalign}
over the bounded state constraint admissible set, $\Xcal.$ The primary goal of this paper is to synthesize an admissible subset, $\Scal\subseteq\Xcal$, 
that is positively invariant under the dynamics \eqref{eq:dynSys} using deterministically sampled data set consisting of successive state pairs, i.e. $\{\x_i, \x_i^+\}_{i=0}^N.$ The set $\Scal$ is an invariant approximation of the maximal \ac{pi} set. 
\begin{defn}\label{def:invSet} \textit{Invariant Set} (\cite{alberto2007invariance}): 
    $\Scal$ is \ac{pi} under the dynamics of \eqref{eq:dynSys} if, $ \forall\x_0 \in \Scal, k \in\mathbb{Z}_{0}^{\infty}$, $T^k(\x_0) \in \Scal.$
\end{defn}
Our secondary goal is to use this data set to synthesize a Lyapunov function over the invariant set, $\Scal$, to verify asymptotic stability the system, while adhering to the constraints. We make the following assumption:
\begin{assum}\label{asmpt:Lipschitz}
    Let \eqref{eq:dynSys} be Lipschitz continuous for some norm, $\norm{\cdot}.$ Let $L>0$ be an upper bound on the Lipschitz constant of the system.
\end{assum}
\assumeref{asmpt:Lipschitz} allows us to extend 
information provided by the samples $\{\x_i,\x_i^+\}_{i=0}^N$ to unsampled states within $\Xcal$. The Lipschitz constant can be determined from data (\cite{wood1996estimation, nejati2023data}).

\subsection{Geometric Conditions of Invariance}
The behavior of subsets of $\Xcal$ under the dynamics of \eqref{eq:dynSys} are characterized geometrically using the precursor and successor sets, defined below.
\begin{defn} \textit{Precursor Set} (\cite{borrelli2017predictive}):
\label{def:precursor}
	For $T: \mathbb{R}^n \rightarrow \mathbb{R}^n,$ the precursor set to set $\Scal$ is $\Pre(\Scal) =\{\x\vecdim{n}\mid T(\x) \in \Scal\}.$
\end{defn}
\begin{defn} \textit{Successor Set} (\cite{borrelli2017predictive}):
\label{def:successor}
	For $T: \mathbb{R}^n \rightarrow \mathbb{R}^n,$ the successor set of set $\Scal$ is $\Suc(\Scal) = \{\x \vecdim{n} \mid \exists \x_0\in \Scal \text{ s.t. } \x = T(\x_0)\}.$ 
\end{defn}
\defrefs \ref{def:precursor} and \ref{def:successor} are leveraged in \lemref \ref{lem:invSet} to determine the geometric conditions of an invariant set, $\Scal$, in $\Xcal$.
\begin{lem} \textit{Invariant Set} (\cite{alberto2007invariance,dorea1999b}): \label{lem:invSet}
    The set $\Scal \subseteq \Omega$ is \ac{pi} for mapping $T: \mathbb{R}^n {\rightarrow} \mathbb{R}^n,$ if $\Suc(\Scal)\subseteq \Scal$ or $\Scal \subseteq \Pre(\Scal)$.
\end{lem}
Methods that use \lemref \ref{lem:invSet} often require an exact model of the system to determine an invariant set (\cite{kerrigan2000robust}). When the system is unmodeled, the invariance condition can only be verified at the data points. With partial information, \cite{strong2025} used Lipschitz continuity to bound the precursor and successor sets.

\begin{lem}\label{lem:sucPrec} (\cite{strong2025}):
    Let \asmptref \ref{asmpt:Lipschitz} hold. Consider the point $\x {\in} \Xcal$ and its successor, $\x^+ {=}T(\x).$  Define $B^+_r(\x) {\defeq }\{\y\vecdim{n} \mid \norm{\x^+ {-} \y}{\leq} L r\}$ and let $B_{r}(\x),B_r^+(\x)\\$$ \subseteq \Xcal$. Then, $ \Suc(B_r(\x)) {\subseteq} B^+_r(\x),$ and $B_r(\x) {\subseteq} \Pre(B^+_r(\x)).$
\end{lem}
\lemref \ref{lem:sucPrec} over approximates the precursor and successor sets of \eqref{eq:dynSys} in $\Xcal$ using Lipschitz continuity rather than requiring full knowledge of \eqref{eq:dynSys}. \lemref \ref{lem:pointsInv} uses these over-approximations to construct an invariant set composed of the union of norm balls about sampled points.
\begin{lem}\label{lem:pointsInv} (\cite{strong2025}):
    Let \asmptref \ref{asmpt:Lipschitz} hold. Consider a data set of $N+1$ pairs, $\{\x_i, \x^+_i\}_{i=0}^N$, sampled in $\Xcal {\subset} \mathbb{R}^n$ where each element in a pair are related via \eqref{eq:dynSys}. Let $B_{r_i}(\x)$ and $B^+_{r_i}(\x_i)$ be as defined in \lemref \ref{lem:sucPrec}.
    If $\cup_{i=0}^N B^+_{r_i}(\x_i) \subseteq \cup_{i=0}^NB_{r_i}(\x_i),$ then  $\cup_{i=0}^NB_{r_i}(\x_i)$ is a \ac{pi} set. 
\end{lem}

\subsection{Lyapunov Functions}

Synthesizing a Lyapunov function in a region of the state space confirms local stability characteristics of a system. Crucially, discrete-time Lyapunov functions only need to be continuous at the equilibrium (\cite{lazar2006model}). \thmref \ref{thm:lyap} states conditions for an equilibrium to be locally asymptotically stable within an invariant set.

\begin{thm} \label{thm:lyap}(\cite{lazar2006model}):
    Let $\Scal \subseteq\mathbb{R}^n$ be a bounded positively invariant set for the system \eqref{eq:dynSys} that contains a neighborhood $\Ncal$ of the equilibrium $\x_e,$ where $T(\x_e) {=} \x_e$. Let $\alpha_1,\alpha_2,\alpha_2 \in \Kcal$.
    Suppose there exists a function $V:\Xcal {\rightarrow} \mathbb{R}_0^{\infty}$ with $V(\x_e){=} 0$ such that
    \begin{subequations}
    \begin{flalign}\label{eq:Lyap1}
        & V(\x) \geq \alpha_1(\norm{\x - \x_e}), \quad \forall \x \in \Scal,
    \end{flalign}
    \begin{flalign}\label{eq:Lyap2}
        & V(\x) \leq \alpha_2(\norm{\x-\x_e}), \quad \forall \x \in \Ncal,
    \end{flalign}
    \begin{flalign}\label{eq:Lyap3}
        & V(T(\x)) - V(\x) \leq -\alpha_3(\norm{\x-\x_e}), \quad \forall \x \in \Scal.
    \end{flalign}
    \end{subequations}
    Then, the equilibrium of \eqref{eq:dynSys} is asymptotically stable in $\Scal.$
\end{thm}

Asymptotic stability can also be found in relation to a set, known as \ac{uub}, which can be confirmed via synthesis of a Lyapunov function (\cite{blanchini2008setTheo}).

\section{Data Driven Invariant Set Synthesis}\label{sec:invSet}

This section presents a data driven algorithm that iteratively computes the intersection of a candidate invariant set $\hat{\Scal}\subseteq \Xcal$ with an over-approximation of its successor ($\hat{\Scal}^+ \supseteq \Suc(\hat{\Scal})$), i.e. $\hat{\Scal} \cap \hat{\Scal}^+$, 
and prunes regions of $\hat{\Scal}$ beyond this intersection to find a true invariant set, $\Scal.$ \alref \ref{alg:findSet} harnesses the approach of the seminal geometric algorithm (\cite{kerrigan2000robust}), but uses a novel data driven approach. We also build on the partitioning strategy of \ac{gaio} (\cite{dellnitz2001algorithms}), but are able to provide true invariance guarantees in finite time and with a finite data set. Invariance is accomplished by parameterizing $\hat{\Scal}$ as the union of max norm balls about sampled points, $\cup_{i=0}^{N} B_{r_i,\infty}(\x_i)$, and $\hat{\Scal}^+$ as $\cup_{i=0}^{N} B_{r_i,\infty}^+(\x_i^+)$, as defined by \lemref \ref{lem:sucPrec}. The result of \alref \ref{alg:findSet} is a set that satisfies $\hat{\Scal} \subseteq \hat{\Scal}^+$. Thus, by \lemref \ref{lem:pointsInv}, $\hat{\Scal}$ is invariant.

\alref \ref{alg:findSet} uses a tree data structure, defined below.
\begin{defn} \label{def:Q}
    Define $\Qcal {=} \Q(\x_i, \x_i^+, r_i, s_i)_{i=0}^N$ as a tree data structure containing $N{+}1$ nodes, $\Q(\x_i, \x_i^+, r_i, s_i)$. Each node contains a state sample, $\x_i$, and the sampled state's successor, $\x_i^+,$ found by applying \eqref{eq:dynSys}. The value $r_i$ defines the radius of the max norm ball about $\x_i$ and is used to construct $B_{r_i,\infty}(\x_i){\subset} \Xcal,$ which partitions $\Xcal.$ The label $s_i$ denotes  if $B_{r_i,\infty}(\x_i)$ is included in ($s_i{=}1$) or excluded from ($s_i{=}0$) the candidate invariant set, $\hat{\Scal}$. Let $L_\Qcal$ denote the set of indices of the $\bar{N}{+}1$ leaf nodes of $\Qcal$ with value $s = 1$.
\end{defn}
We define $\hat{\Scal}$ using the leaf nodes of $\Qcal$ where $s_i=1$
\begin{defn}\label{def:hatS}
    Define $\hat{\Scal} {=}  \cup_{k\in L_\Qcal} B_{r_k,\infty}(\x_k),$ where each norm ball is constructed from the leaf nodes of $\Qcal$ where $s=1.$ For brevity, we denote $\Qcal_{\hat{\Scal}}=\Q(\x_i, \x_i^+, r_i, s_i)_{i\in L_\Qcal}.$
\end{defn}

\alref \ref{alg:findSet} iteratively determines an invariant set as follows. \alref \ref{alg:findSet} initializes $\hat{\Scal}$ with a partition of $\Xcal$ (or an under-approximation of $\Xcal$) using $\cup_{i \in L_\Qcal} B_{r_i,\infty}(\x_i) \subseteq \Xcal$ constructed from $\Qcal.$ Each node of the initial partition is part of the candidate invariant set. 
The lower threshold for the radius of a node, $\tau > 0$, and an upper bound on the system's Lipschitz constant with respect to the max norm, $L>0$, are assumed given for initializing the algorithm. Note a lower threshold, $\tau$, allows for larger data sets.

%

\begin{figure}
\begin{subfigure}[t]{0.3\columnwidth}
    \centering
     \includegraphics[width=\textwidth]{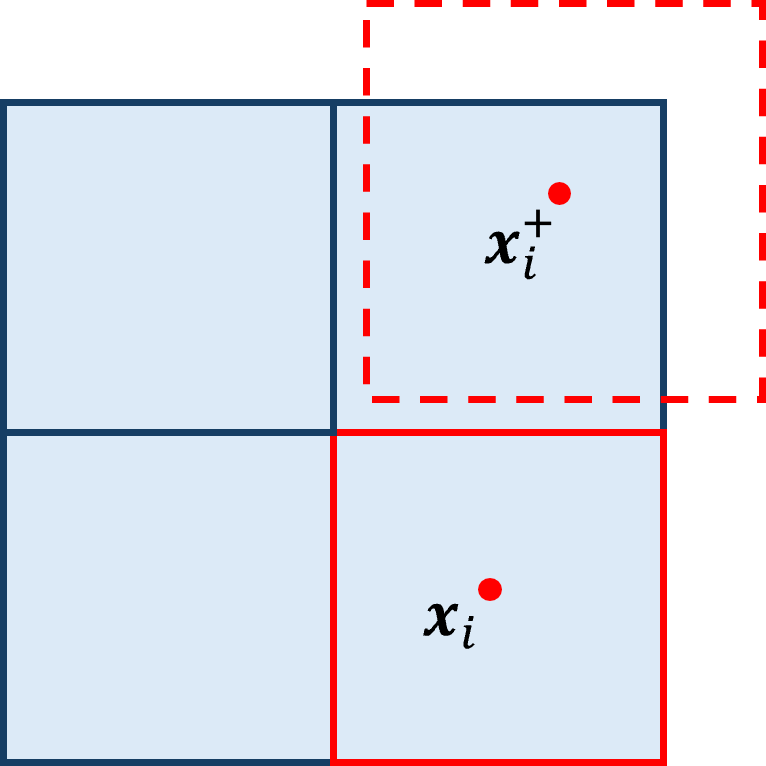}
     \caption{The candidate invariant set intersected with the over-approximation of a node's successor set.}
     \label{fig:algEx1}
\end{subfigure}
\hfill
\begin{subfigure}[t]{0.3\columnwidth}
    \centering
     \includegraphics[width=\textwidth]{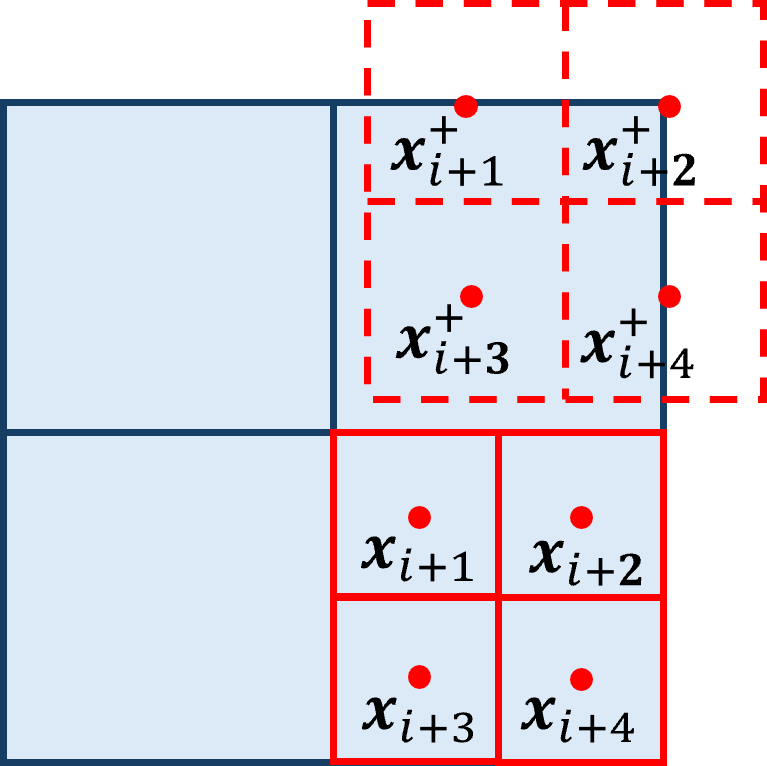}
     \caption{The resulting division of a node (\alref \ref{alg:divide}) and its corresponding samples.}
     \label{fig:algEx2}
\end{subfigure} 
\hfill
\begin{subfigure}[t]{0.3\columnwidth}
    \centering
     \includegraphics[width=0.875\textwidth]{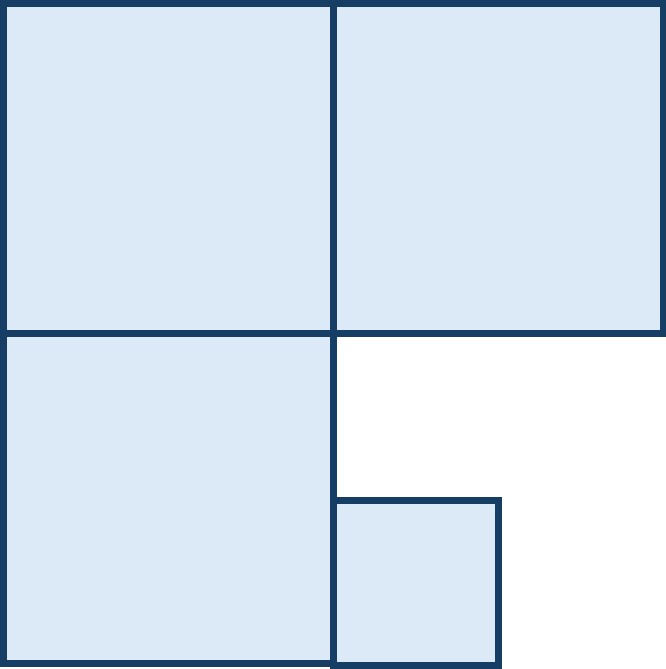}
     \caption{The new candidate invariant set resulting from the steps in \ref{fig:algEx1} and \ref{fig:algEx2}.}
     \label{fig:algEx3}
\end{subfigure} 
\caption{A representation of different stages in \alref \ref{alg:findSet} for a single node in the tree data structure. Each blue square is a member of the candidate invariant set, $\hat{\Scal}.$}
\label{fig:algEx}
\end{figure}
At each iteration of \alref \ref{alg:findSet}, the behavior of \eqref{eq:dynSys} in $\hat{\Scal}$ is characterized by constructing $B^+_{r_i,\infty}(\x_i)$ for each node of $\Qcal_{\hat{\Scal}}$ using \lemref \ref{lem:sucPrec}, $L$, and the successive sample $\x_i^+$. Here, the tree structure, $\Qcal$, is essential for efficiently characterizing the behavior of $B^+_{r_i,\infty}(\x_i)$. If $B^+_{r_i,\infty}(\x_i)$ fully remains within $\hat{\Scal}$, then the corresponding node is unchanged, i.e. $s_i$ remains $1$ for node $\Q(\x_i, \x_i^+, r_i, s_i)$. If $B^+_{r_i,\infty}(\x_i)$ does not intersect with $\hat{\Scal}$ at all, it is removed from the candidate invariant set for all future iterations, i.e. $s_i=0$. 
If $B^+_{r_i,\infty}(\x_i)$ only partially remains within $\hat{\Scal}$, there are two possible outcomes. If $\frac{r_i}{2}{<}\tau$, then the radius is too small and the node is removed from the candidate invariant set for all future iterations, i.e. $s_i=0$. 
In all above cases, $\Q(\x_i, \x_i^+, r_i, s_i)$ remains a leaf node.
If $\frac{r_i}{2}\geq\tau$, then \alref \ref{alg:divide} divides the node into $2^n$ new leaf nodes (therefore creating $2^n$ new samples) with new radii of $\frac{r_i}{2}$. These new nodes are added to $\Qcal$ and remain  the candidate invariant set by setting $s_i=1$. 
Each iteration, $L_\Qcal$ and $\hat{\Scal} =\cup_{k\in L_\Qcal} B_{r_k,\infty}(\x_k)$ update as each node in $\Qcal_{\hat{\Scal}}$ is considered. The dataset $\Dcal$ is tracks the samples that make up $\hat{\Scal}$ each iteration. 
\figref \ref{fig:algEx} shows an example of \alref \ref{alg:findSet} applied to a single node in $\Xcal \subset \IR^2$ for \eqref{eq:dynSys}.

\alref \ref{alg:findSet} terminates when the candidate invariant set remains the same after iterating through each node of $\bar{Q}(\x_i, \x_i^+,r_i,1)_{i=0}^{\bar{N}}$. \thmref \ref{thm:alg} shows \alref \ref{alg:findSet} results in an invariant set. The radius threshold $\tau$ ensures termination in finite time by precluding infinite divisions.

\begin{thm}\label{thm:alg}
    Let \asmptref \ref{asmpt:Lipschitz} hold for \eqref{eq:dynSys}, where $L$ is defined with respect to the max-norm. Let $\Qcal$ (\defref \ref{def:Q}) $\tau {>}0$, and $L{>}0$ be inputs to  \alref \ref{alg:findSet}. \alref \ref{alg:findSet} will produce an invariant set, $\Scal$, in a finite number of steps.
\end{thm}
\begin{pf}
    \alref \ref{alg:findSet} terminates when $\tilde{\Scal}_{j-1}{=} \hat{\Scal}_{j-1}$. If $\tilde{\Scal}_{j-1} {=} \hat{\Scal}_{j-1} {\neq} \emptyset$ at termination, then by lines 7 through 17, there were no node divisions or node removals from $\hat{\Scal}_{j-1}$ in iteration $j-1$, i.e. every $B_{r_i, \infty}(\x_i)$ in $\Scal {=} \cup_{k \in L_\Qcal} B_{r_k,\infty}(\x_k)$ 
    had a corresponding over approximation of its successor set, $B^+_{r_i,\infty}(\x_i)$ that was a subset of $\Scal$. In other words, $\cup_{k \in L_\Qcal} \Suc(B_{r_i, \infty}(\x_i)) {\subseteq} \cup_{k \in L_\Qcal} B^+_{r_i, \infty}(\x_i){ \subseteq} \cup_{k \in L_\Qcal} B_{r_i, \infty}(\x_i){ =} \Scal.$ By \lemref \ref{lem:pointsInv}, $\Scal$ is an invariant set.

    Because divisions of the partition of $\Xcal$ are limited by the radius threshold, $\tau,$ there are a finite number of divisions that can be performed. After that, each time there is a node where $B^+_{r_i,\infty}(\x_i) {\not\subseteq} \hat{\Scal}_j$, 
    the node $Q(\x_i,\x_i^+,r_i,s_i)$ is removed from $\hat{\Scal}_j$. This process iterates until $\tilde{\Scal}_{j-1} {=}\hat{\Scal}_{j-1},$ which may only occur at the null set, which is invariant.

    The number of divisions and nodes of $\Qcal$ are limited by $\tau$. Thus, \alref \ref{alg:findSet} terminates in finite time. \hfill $\square$

\end{pf}

\begin{algorithm}
    \caption{Synthesize Invariant Set}
    \label{alg:findSet}
    \begin{algorithmic}[1]
        \Require $\Qcal = \Q(\x_i, \x_i^+, r_i, s_i=1)_{i=0}^{N}$, $\tau>0$, $L > 0$
        \State  $j = 0$, $\hat{\Scal}_j \leftarrow \cup_{k \in L_\Qcal B_{r_k,\infty}(\x_k)}$, $\tilde{\Scal}_{j-1} \leftarrow \emptyset$
        \While{$\hat{\Scal}_{j-1} \neq \tilde{\Scal}_{j-1}$}
        \State $\hat{\Scal}_j \leftarrow \cup_{k \in L_\Qcal} B_{r_k,\infty}(\x_k)$, $\tilde{\Scal}_j \leftarrow \cup_{k \in L_\Qcal} B_{r_k,\infty}(\x_k)$
        \State $\Dcal = \{\}$
        \For{$\Q(\x_i, \x_i^+, r_i, s_i) \in  \Qcal_{\tilde{\Scal}_j}$}
        \State $B^+_{r_i,\infty}(\x_i) \defeq \{\y\vecdim{n}\mid\norm{\x_i^+-\y}_{\infty}\leq Lr_i\}$
        \If{$B^+_{r_i,\infty}(\x_i) \cap \hat{\Scal}_j = B^+_{r_i,\infty}(\x_i)$}
        \State $s_i=1$ in $\Q(\x_i, \x_i^+,r_i, s_i)$
        \State $\Dcal = \Dcal \cup (\x_i, \x_i^+)$
        \ElsIf{$B^+_{r_i,\infty}(\x_i) \cap \hat{\Scal}_j = \emptyset$}
        \State  $s_i=0$ in $\Q(\x_i, \x_i^+,r_i, s_i)$
        \Else
        \If{$\frac{r_i}{2} >\tau$}
        \State  $\Qcal$ = Alg. \ref{alg:divide}$(\Qcal, i)$
        \Else
        \State $s_i=0$ in $\Q(\x_i, \x_i^+,r_i, s_i)$
        \EndIf
        \EndIf
        \State $\hat{\Scal}_j \leftarrow \cup_{k \in L_\Qcal} B_{r_k,\infty}(\x_k)$
        \EndFor
        \State $j =j +1$
        \EndWhile
        \State $\Scal \leftarrow \hat{\Scal}_j$
    \end{algorithmic}
    \Return $\Scal$, $\Dcal$
\end{algorithm}

\begin{algorithm}
    \caption{Divide Node $i$}
    \label{alg:divide}
    \begin{algorithmic}[1]
        \Require $\Qcal= \Q(\x_i, \x_i^+, r_i, s_i)_{i=0}^{N}$, $i$
        \For{$k = N+1$ to $k = N + 1+ 2^n$}
        \State Sample $\x_k = \x_i + \frac{r_i}{2}{\mathbf{v}(k-N)}$, where $\mathbf{v}(k-N)\in \{1, -1\}^n$, and sample $\x_k^+$ by applying \eqref{eq:dynSys}
        \State Add leaf node $\Q(\x_k, \x_k^+,r_k, s_k)$ to node i
        \State $s_k = 1$ in $\Q(\x_k, \x_k^+,r_k, s_k)$
        \EndFor 
    \end{algorithmic}
    \Return $\Qcal = \Q(\x_i, \x_i^+,r_i, s_i)_{i=0}^{N+k}$
\end{algorithm}

\alref \ref{alg:findSet} differs from \ac{gaio} (\cite{dellnitz2001algorithms}) in that it uses the system's Lipschitz continuity to over approximate the successor set of a max norm ball about a single sample, characterizing all states in a partition with a single $\{\x,\x^+\}$ pair. Further, \algoref{alg:findSet} is guaranteed to terminate in finite time (with finite data) with a true invariant set rather than a covering of the set.

Lipschitz continuity is a conservative characterization of a system -- introducing conservatism into \alref \ref{alg:findSet}. To find a viable invariant set $\Scal \subseteq \Xcal$, the dynamics of \eqref{eq:dynSys} must ensure $B_{\frac{\tau}{2},\infty}^+(\x_k) \subseteq \Scal$ for all $k\in L_\Qcal$. If the dynamics make this impossible, \alref \ref{alg:findSet} produces an empty set. Note also that \alref \ref{alg:findSet} suffers from the curse of dimensionality, as partitioning a node creates $2^n$ new nodes (\cite{meagher1982geometric}).

\section{Data Driven Lyapunov Function Synthesis}

The aim of this section is to synthesize a Lyapunov function, $V:\Scal \rightarrow \mathbb{R}$, using the invariant approximation of the maximal \ac{pi} set from \algoref{alg:findSet}, $\Scal$, and its corresponding samples, $\Dcal$, to confirm \ac{uub}. As in \secref \ref{sec:invSet}, the main tool used is max-norm balls about data points. 
We consider the case where $V$ is piecewise constant function -- defined by a constant, $v_i$, on each max norm ball of $\Scal {=} \cup_{k \in L_\Qcal} B_{r_k,\infty}(\x_k)$ -- and derive conditions for the value $v_i$ on each partition so that $V$ shows asymptotic convergence of \eqref{eq:dynSys} to an invariant approximation of the minimal \ac{pi} set, $\Scal_\beta.$ The set $\Scal_\beta$ contains a
user defined set $\Bcal_\Ecal$, where the system's energy may increase. These conditions can be used to synthesize a Lyapunov function via optimization.

There are two main challenges in synthesizing $V$. The first challenge is that the decrease condition of the Lyapunov function must be enforced across the entirety of $\Scal.$ We exploit the unique tree structure of $\Scal$ and $V$ to develop a decrease condition that can be enforced by a single condition on a node, but ensures the decrease condition holds across the node. To reference the tree structure in upcoming proofs, we assume the following.
\begin{assum}\label{asmpt:SQ}
 Let $\Qcal_\Scal$, $\Scal$, and $\Dcal =\{(\x_i,\x_i^+)\}_{i \in L_\Qcal}$, and be the tree nodes, invariant set (\defref \ref{def:hatS}), and data set produced by applying \algoref{alg:findSet} to $\Xcal$, while sampling state successors from \eqref{eq:dynSys}, which satisfies \assumeref{asmpt:Lipschitz} with respect to the max norm.
\end{assum}

The second challenge involves the system's equilibrium ($\x_e$). By \thmref \ref{thm:lyap}, $V(\x_e) {=} 0$ and, accordingly, $V(T(\x_e)) - V(\x_e) {=} 0.$ The node in $\Qcal_\Scal$ containing the equilibrium cannot be constant without either violating the positive definiteness of $V$ or the condition $V(\x_e) {=} 0.$ 
Therefore, we leverage \ac{uub} (\cite{blanchini2008setTheo}). We consider a user defined set $\Bcal_\Ecal {\subset} \Scal$ where the decrease condition need not hold and develop conditions for $V$ to verify that any $\x {\in} \Scal{\setminus}\Scal_\beta$ asymptotically converges to the invariant approximation of the minimal \ac{pi} set, $\Scal_\beta \supseteq \Bcal_\Ecal$.




\subsection{Convergence to a set}\label{sec:set}
%
This section analyzes a system's convergence to $\Scal_\beta \supseteq\Bcal_\Ecal$, an invariant estimate of the minimal \ac{pi} set. To this end, we define a Lyapunov-like function form, the values of which are later selected via optimization.
\begin{defn}\label{def:V}
    Let \assumeref{asmpt:SQ} hold and let $\Bcal_\Ecal {\subset} \Scal$. Define $V{:}\Scal {\rightarrow} \mathbb{R}$ as a discontinuous piecewise constant function described by $v_{\x_i}$ on each node $i {\in} L_\Qcal$ in $\Qcal_\Scal.$ For any $\z{\in} \Scal,$ 
    \begin{flalign}\label{eq:vDef}
    V(\z) = 
        \begin{cases}
            & v_{\x_i},\quad \quad  \quad \z \in B^{\circ}_{r_i,\infty}(\x_i)\setminus \Bcal_\Ecal\\
            & \min_{j \in \mathbb{Z}_1^e}v_{\x_j}, \quad \z \in \cap_{j=1}^{e} \partial B_{r_j,\infty}(\x_j) \\
            & v_{\x_i}=0, \quad \z \in \Bcal_\Ecal,
        \end{cases}
    \end{flalign}
    where $\cap_{j=1}^{e} \partial B_{r_j,\infty}(\x_j)$ describes the intersection of $e$ neighboring nodes where $\z$ lies on the boundary.
\end{defn}

We develop a decrease condition for each node of $\Scal$ to ensure \condref \eqref{eq:Lyap3} holds for $\Scal {\setminus} \Bcal_\Ecal$ before finding conditions on $V$ for \eqref{eq:dynSys} in $\Scal{\setminus}\Scal_\beta$ to converge to $\Scal_\beta$.

\subsubsection{Decrease Condition}
The decrease condition of the Lyapunov function (\condref \eqref{eq:Lyap3}) must be enforced on all $\x{\in}\Scal{\setminus}\Bcal_\Ecal$. However, the available data is limited to $\{\x,\x^+\}_{i=0}^{\bar{N}}$. 
We leverage \lemref \ref{lem:sucPrec} and the unique format of $V$ to develop an inequality that, when enforced at a data point ($\x_i$) contained in a node of the tree data structure, ensures the decrease condition holds across $B_{r_i,\infty}(\x_i)$.

\begin{lem}\label{lem:vBound}
    Let \assumeref{asmpt:SQ} hold. Let $V$ be defined by \defref \ref{def:V}. Let $\hat{B}_{r_i,\infty}(\x_i) {=} \{\z {\in} B_{r_i,\infty}(\x_i) {\mid} V(\z) {=}V(\x_i)\}.$
    Let $\hat{\alpha} \in \Kcal,$ $-\underline{\hat{\alpha}}_{i} \leq \min_{\y\in B_{r_i,\infty}(\x_i)}{-}\hat{\alpha}(\Delta(\y, \Bcal_\Ecal))$, and \\ $\bar{V}_i^+ {\defeq} \max_{\y \in B^+_{r_i,\infty}(\x_i)} V(\y).$
    If
    \begin{flalign}\label{eq:lemDecrease1}
        & \bar{V}_i^+ - V(\x_i) \leq -\underline{\hat{\alpha}}_{i},
    \end{flalign}
    then $V(\z^+) {-} V(\z) {\leq} {-}\hat{\alpha}(\Delta(\z, \Bcal_\Ecal))$ for all $\z {\in} \hat{B}_{r_i,\infty}(\x_i).$ 
\end{lem}
\begin{pf}
    From \lemref \ref{lem:sucPrec}, $\Suc(B_{r_i,\infty(\x_i)}) {\subseteq} B^+_{r_i,\infty}(\x_i).$ Therefore, $\bar{V}_i^+ {\geq} \max_{\z{\in}{B_{r_i,\infty(\x_i)}}} V(\z^+).$ Because $V(\z)$ is constant for all $\z{\in} \hat{B}_{r_i,\infty}(\x_i),$ $V(\z^+) {-} V(\z) {\leq} \bar{V}_i^+ {-} V(\x_i) \leq {-}\underline{\hat{\alpha}}_{i} {\leq} {-}\hat{\alpha}(\Delta(\z, \Bcal_\Ecal))$ holds across $\hat{B}_{r_i,\infty}(\x_i)$. Hence, \eqref{eq:lemDecrease1} implies $V(\z^+) {-} V(\z) {\leq} {-}\hat{\alpha}(\Delta(\z, \Bcal_\Ecal))$ holds $\forall \z {\in} \hat{B}_{r_i,\infty}(\x_i).$ \hfill $\square$
\end{pf}
The practical utility of \lemref \ref{lem:vBound} depends on the tree structure of $\Scal$ and $V.$ Typically, determining the value of $\bar{V}_i^+$ for $B_{r_i,\infty}(\x_i)$ would require additional sampling or the system model. Here, it requires determining all nodes in $\Qcal_\Scal$ that $B^+_{r_i,\infty}(\x_i)$ intersects with by traversing $\Qcal$ and then determining the values of $V$ based on \eqref{eq:vDef}. 

\subsubsection{Lyapunov Function Conditions}

In \thmref \ref{thm:set}, conditions on $V:\Scal {\rightarrow} \mathbb{R}$ (\defref \ref{def:V}) are found to verify asymptotic convergence of \eqref{eq:dynSys} to $\Scal_{\beta}$, a sublevel set of $V$ which contains the user defined set $\Bcal_\Ecal.$

\begin{thm}\label{thm:set}
    Consider \eqref{eq:dynSys}. Let \assumeref{asmpt:SQ} hold, and let $V:\Scal\rightarrow\mathbb{R}$ be defined by \defref \ref{def:V}, where $\Bcal_\Ecal \defeq \cup_{k=m}^p B_{r_k,\infty}(\x_k) \subset \Scal$, $m \leq p$, and  $\mathbb{Z}_m^p\subseteq L_\Qcal$. Let $\hat{\alpha} \in \Kcal.$ 
    Let $V$ satisfy
    \begin{subequations} 
    \begin{flalign}
        & v_{\x_i} \leq 1, \quad \forall i\in L_\Qcal \label{eq:maxV_thm2}\\ 
        & v_{\x_i} > 0, \quad \forall i\in L_\Qcal\setminus\mathbb{Z}_m^p, \label{eq:positiveDef_thm2}\\ 
        \begin{split}\label{eq:v1Decrease_thm2}
            & \bar{V}_i^+ - V(\x_i) \leq -\underline{\hat{\alpha}}_i, \quad \forall i \in L_\Qcal\setminus\mathbb{Z}_m^p, 
        \end{split} \\
        &  \bar{V}_\Ecal^+ \leq \beta \label{eq:Vincrease} 
    \end{flalign}
    \end{subequations}
    where $\beta \in \mathbb{R}_{0}^{\underline{1}},$ $-\underline{\hat{\alpha}}_i {\leq} \min_{\y\in B_{r_i,\infty}(\x_i)}-\hat{\alpha}(\Delta(\y,\Bcal_\Ecal))$, $\bar{V}_i^+ \defeq \max_{\y \in B^+_{r_i,\infty}(\x_i)} V(\y),$ and $\bar{V}_\Ecal^+ \defeq \max_{\y \in \cup_{k=m}^p B^+_{r_i,\infty}(\x_k)} V(\y)$. Then, \eqref{eq:dynSys} asymptotically converges to $\Scal_\beta \supseteq \Bcal_\Ecal,$ where $\Scal_\beta \defeq \{\x\in\Scal \mid V(\x) \leq \beta\}$ and $\Scal_\beta$ is an invariant set.
\end{thm}
\begin{pf}
    By \eqref{eq:maxV_thm2} and \eqref{eq:positiveDef_thm2}, $V$ is bounded and positive definite on $\Scal{\setminus}\Bcal_\Ecal.$ By \defref \ref{def:V}, $V$ is bounded on $\Bcal_\Ecal.$
    
    Let $\hat{B}_{r_i,\infty}(\x_i) {=} \{\z {\in} B_{r_i,\infty}(\x_i) {\mid} V(\z) {=} V(\x_i)\}.$ By \defref \ref{def:V}, $\cup_{i \in L_\Qcal\setminus[m,...,p]}\hat{B}_{r_i,\infty}(\x_i)$
    covers $\Scal{\setminus}\Bcal_\Ecal.$ By \eqref{eq:v1Decrease_thm2} and \lemref \ref{lem:vBound}, the Lyapunov decrease condition holds on all $\hat{B}_{r_i,\infty}(\x_i) \subseteq \Scal\setminus\Bcal_\Ecal$ and therefore on all of $\Scal{\setminus}\Bcal_\Ecal.$ 

    Paralleling the proof of \thmref 2.2.4 (\cite{lazar2006model}), if $\z {\in} \Scal{\setminus}\Bcal_\Ecal$, then $\left( V(T^{j+1}(\z)) - V(T^j(\z)) \right)\xrightarrow{j\rightarrow T} 0$ for some $T{>}0$ by \condrefs \eqref{eq:positiveDef_thm2} and \eqref{eq:v1Decrease_thm2}. Since this bounds $-\hat{\alpha}(\Delta(\z,\Bcal_\Ecal))\leq 0$ below, $\hat{\alpha}(\Delta(T^j(\z), \Bcal_\Ecal)) \xrightarrow{j\rightarrow T} 0.$ Therefore, $\Delta(T^j(\z),\Bcal_\Ecal) \xrightarrow{j\rightarrow T} 0$ for all $\z \in \Scal\setminus\Bcal_\Ecal.$
    
    By Conditions~\eqref{eq:Vincrease} and \eqref{eq:v1Decrease_thm2} respectively, the successors of $\z{\in}\Bcal_\Ecal{\subseteq}\Scal_\beta$ and $\z{\in}\Scal_\beta{\setminus}\Bcal_\Ecal$ will be in $\Scal_\beta.$ Thus, $\Scal_\beta$ is invariant and all trajectories starting in $\Scal\setminus\Scal_\beta$ converge to it.\hfill $\square$
    

\end{pf}

\begin{cor}
    If a Lyapunov function satisfying \thmref \ref{thm:set} for \eqref{eq:dynSys} in $\Scal$ is found, then \eqref{eq:dynSys} is uniformly ultimately bounded in $\Scal.$
\end{cor}

\section{Numerical Examples}

\subsection{Linear Dynamical System}\label{sec:numExLin}
\begin{figure}
    \centering
    \includegraphics[width=0.8\linewidth]{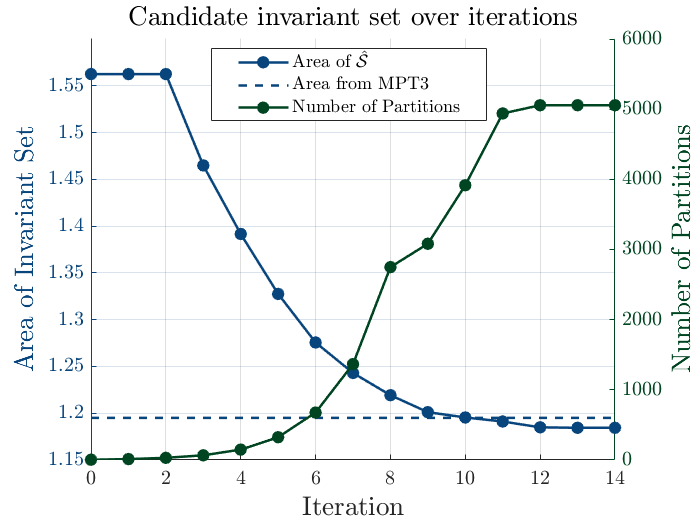}
    \caption{Area and number of partitions of the candidate invariant set over each iteration of \alref \ref{alg:findSet}. The area of $\hat{\Scal}$ is compared to the maximal invariant set area calculated from MPT3 (\cite{MPT3}).}
    \label{fig:invSetStats_linear}
\end{figure}

Consider the linear system
\begin{flalign}\label{eq:exLinSys}
    \x^+ =\bmat{0.2200 & 0.4013 \\ -0.5364 & 0.2109}\x,
\end{flalign}
where $\Xcal: [-0.25,1]\times[-1, 0.25]$. An upper bound on the Lipschitz constant with respect to the max norm is $L {=} 0.8225.$ \alref \ref{alg:findSet} was initialized with $\hat{\Scal} {=} \Xcal$ and $\tau {=} 0.001$. 
The candidate invariant set area is compared to the area of the maximal invariant set found using MPT3 in \figref \ref{fig:invSetStats_linear} --showing the invariant set produced by \alref \ref{alg:findSet} has a slightly smaller area than that found by MPT3. The number of partitions required used was also tracked across iterations of the algorithm (\figref \ref{fig:invSetStats_linear}). The total number of sampled $(\x,\x^+)$ pairs required to produce $\Scal$ was 11,796. The total number of partitions was 5,056. In contrast, \cite{strong2025}, required 9,190 partitions to iteratively expand an initial invariant set to a similar size.


\begin{figure}
    \centering
    \includegraphics[width=0.8\linewidth]{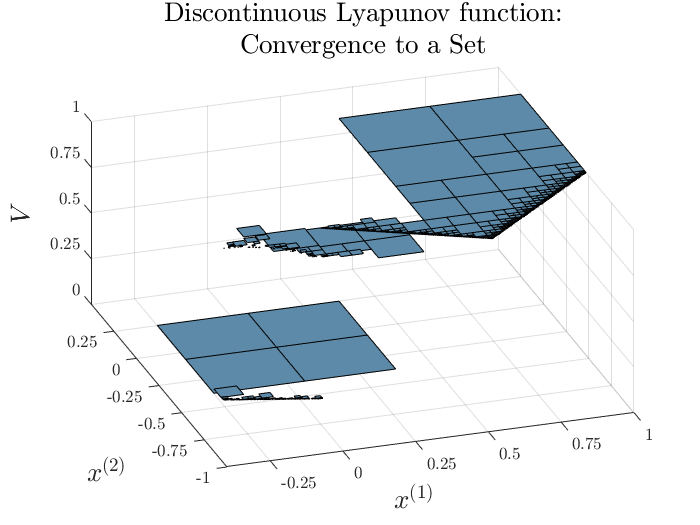}
    \caption{A discontinuous Lyapunov function shows asymptotic convergence of \eqref{eq:exLinSys} to the small invariant set, $\Bcal_\Ecal = [-0.25,0.375]\times [-0.375,0.25]$. }
    \label{fig:lyapSet}
\end{figure}

Next, we sought a scalar function to verify asymptotic convergence of the system. An optimization problem was set up using \thmref \ref{thm:set}, where $\beta$ was minimized. For \condref \eqref{eq:v1Decrease_thm2}, $\underline{\hat{\alpha}}_i{=}c\Delta(\x_i,\Bcal_\Ecal) {+}c\sqrt{2}r_i$ was used, where $c{\geq}0.25$ and  $\x_i$ and $r_i$ are defined by the relevant node in the tree. \figref \ref{fig:lyapSet} shows the Lyapunov function verifying that \eqref{eq:exLinSys} converges to the set $\Bcal_\Ecal {=} [-0.25,0.375]{\times} [-0.375,0.25]$. Here, $\bar{V}^+_\Ecal {\leq} \beta$ holds for $\beta {=}0,$ meaning $\Bcal_\Ecal$ is invariant. 

\subsection{Nonlinear Dynamical System}

\begin{figure}
    \centering
    \includegraphics[width=0.8\linewidth]{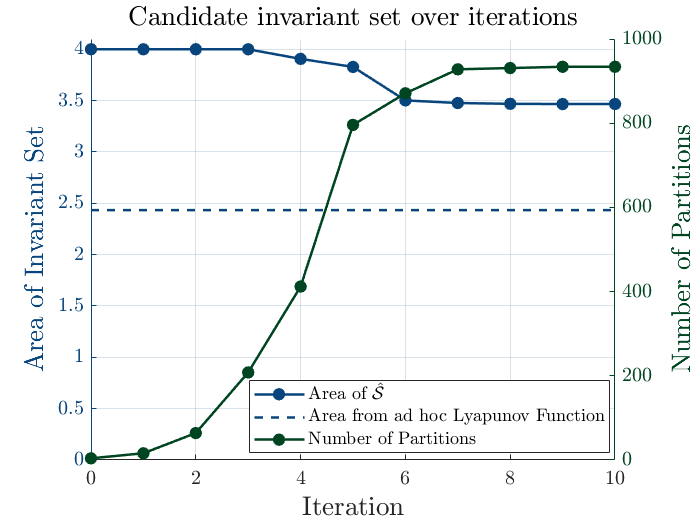}
    \caption{Area and number of partitions of the candidate invariant set over each iteration of \alref \ref{alg:findSet}.}
    \label{fig:invSetStats_nonlinear}
\end{figure}

Consider the nonlinear dynamical system
\begin{flalign}\label{eq:exNonLinSys}
    \begin{split}
        & x_1^+ = 0.5x_1 -0.7x_2^2 \\
        & x_2 = 0.9x_2^3 +x_1x_2
    \end{split}
\end{flalign}
over the space, $\Xcal: [-1, 1]\times [-1, 1],$ with $L = 5.728.$ upper bounding its Lipschitz constant with respect to the max norm. \alref \ref{alg:findSet} with $\tau = 0.01$ and $\hat{\Scal}$ initialized as $\Xcal$ was used to determine an invariant set.
\figref \ref{fig:invSetStats_nonlinear} shows the area of the candidate invariant set over each iteration of the algorithm compared to the area of the invariant set of the level set of the Lyapunov function, $V(\x) = x_1^2 + x_2^2$. The invariant set produced by \alref \ref{alg:findSet} has a larger invariant set than that of the level set and required 934 partitions. In total, 2,178 sampled $(\x,\x^+)$ pairs were needed for \alref \ref{alg:findSet} to determine $\Scal.$


\begin{figure}
    \centering
    \includegraphics[width=0.8\linewidth]{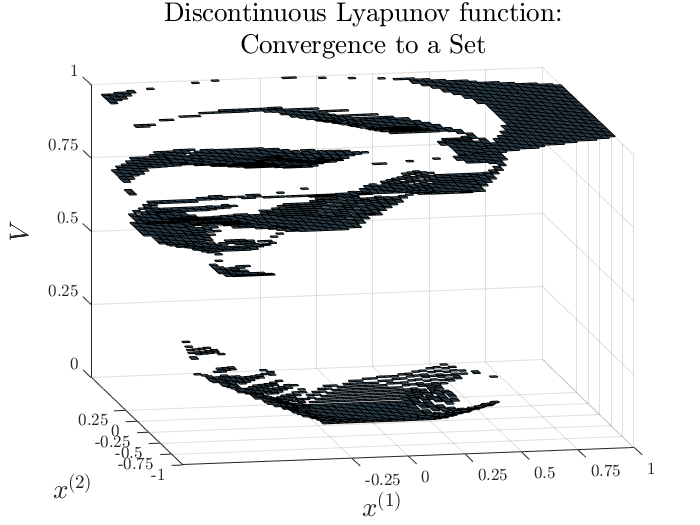}
    \caption{A discontinuous Lyapunov function shows asymptotic convergence of \eqref{eq:exNonLinSys} to the set, $\Bcal_\Ecal = [-0.21875,0.21875]\times [-0.21875, 0.21875]$. }
    \label{fig:lyapSet_nonlinear}
\end{figure}

\figref \ref{fig:lyapSet_nonlinear} shows the scalar function verifying convergence of \eqref{eq:exNonLinSys}, where $\Bcal_\Ecal{ =} [-0.21875,0.21875]\times [-0.21875, 0.21875]$. The optimization problem was created using \thmref \ref{thm:set} where $\beta$ was minimized. Here, $\underline{\hat{\alpha}}_i {=} c\Delta(\x_i,\Bcal_\Ecal) +c\sqrt{2}r_i$ was used with $c{\geq} 0.15$ and $\x_i$ and $r_i$ defined by the node. Additional data was needed to find the Lyapunov function; all nodes were divided until they hit the threshold, $\tau$ -- producing 3,946 pairs of $\{\x,\x^+\}$ data. \condref \eqref{eq:Vincrease} held with $\beta {=} 0.0057$ -- thus, the sublevel set $V(\x) \leq 0.0057$ is the invariant approximation of the minimal \ac{pi} set.

\bibliography{ifacconf}             
\end{document}